\documentclass[aps,showpacs,prl,twocolumn]{revtex4}%
\usepackage{amsfonts}
\usepackage{amsmath}
\usepackage{amssymb}
\usepackage{graphicx}%

\begin{document}
\title{Zero sound in a single component fermion - Bose Einstein Condensate mixture}
\author{D. H. Santamore}
\affiliation{ITAMP, Harvard-Smithsonian Center for Astrophysics, Cambridge, Massachusetts 02138}
\affiliation{Department of Physics, Harvard University, Cambridge, Massachusetts 02138}
\author{Sergio Gaudio}
\affiliation{Department of Physics, Boston College, Chestnut Hill, Massachusetts 02467}
\affiliation{T-11, Theory division, Los Alamos National Laboratory, Los Alamos, New Mexico 87545}
\author{Eddy Timmermans}
\affiliation{T-4, Theory division, Los Alamos National Laboratory, Los Alamos, New Mexico 87545}

\pacs{05.30.Jp, 03.75.Kk, 32.80.Pj, 67.90.+z}

\begin{abstract}
The resonant dynamics of mediated interactions supports zero-sound in a cold
atom degenerate mixture of a single component fermion gas and a Bose-Einstein
condensate (BEC). We characterize the onset of instability in the phase
separation of an unstable mixture and we find a rich collective mode structure
for stable mixtures with one undamped mode that exhibits an avoided crossing
and a Landau-damped mode that terminates.

\end{abstract}
\date{\today }
\maketitle

Recent successes in cold atom fermion cooling
\cite{GrimJinKetSalThom} have opened up possibilities to study
Fermi-liquid in depth, beyond the realm of fermion superfluidity.
Cold atom fermions are cooled `sympathetically' by using a
Bose-Einstein condensate (BEC) as a refrigerator \cite{EXPT0102}.
The BEC mixture, if it is cooled below $10$\% of the
Fermi-temperature, would result in the first neutral
Fermi-liquid/boson superfluid mixtures to be investigated in the
laboratory \cite{Eddy} since the $^{3}$He--$^{4}$He-studies. Cold
atom technology is ideally suited for revisiting interesting
features of the Fermi-liquid/boson superfluid phase diagram
\cite{Baym} whose microscopic description has proven to be a
considerable challenge in the case of the strongly coupled
$^{3}$He--$^{4}$He systems \cite{KS93}. For instance, a
fermion-boson Feshbach resonance can vary the mediated interactions
that play a crucial role in the phase separation of
$^{3}$He--$^{4}$He \cite{BBP66}.

In this letter, we show that a dilute fermion gas of single spin
atoms immersed in a BEC exhibits a genuine collective Fermi-liquid
behavior by supporting a long-lived zero-sound excitation due to the
BEC-mediated fermion--fermion interactions. The observation of this
mode would present a clear manifestation of mediated interactions
since the Pauli exclusion-principle prevents the neutral fermion
atoms from interacting directly at the experimental densities. It
would also illustrate the importance of the dynamics of the mediated
interactions: In the case of phonon-mediated electron-electron
interactions, the sign of the interactions (overscreening) can be
reversed \cite{Pines}, thus, supporting zero-sound instead of having
only the fermion-fermion attraction predicted by the static
description of the mediated interaction \cite{Stoof}. In addition,
if the Fermi-velocity $v_{\mathrm{F}}$ significantly exceeds the BEC
sound velocity $c$ and the mixture is stable, the system supports a
second collective excitation with a dispersion relation that
terminates at a specific wavenumber. Furthermore, we find that the
imaginary part of the eigenvalue in the dispersion relation does not
necessarily signal the instability of the homogeneous mixture as
claimed by Ref.\ \cite{PZWM02}, but can describe Landau damping.

We consider a quantum degenerate gas mixture of neutral fermions of
mass $m_{ \mathrm{F}}$ and bosons of mass $m_{\mathrm{B}}$,
uniformly distributed at equilibrium densities
$\rho_{\mathrm{F}}^{0}$ and $\rho_{\mathrm{B} }^{0} $, respectively.
The gas is dilute so that the average inter-particle distances
$\left(  \rho_{\mathrm{F}}^{0}\right)  {}^{-1/3}$ for fermions and
$\left( \rho_{\mathrm{B}}^{0}\right)  {}^{-1/3}$ for bosons
significantly exceed the range of the interaction potentials as well
as the scattering lengths $a_{ \mathrm{BB}}$ and $a_{\mathrm{FB}}$
of the binary atom boson-boson and boson-fermion interactions. The
temperature is sufficiently low to validate the mean-field
description of the BEC and the use of the collisionless
approximation in treating the fermion dynamics, but is higher than
the critical temperature for fermion pairing. The fermion and boson
quantum liquids are coupled by the mean-field interaction with the
interaction strength
$\lambda_{\mathrm{FB}}=(2\pi\hbar^{2}a_{\mathrm{FB} })\left(
1/m_{\mathrm{B}}+1/m_{\mathrm{F}}\right)  $. The interaction energy
of fermion or boson is proportional to the local density of the
other liquid: $\lambda_{\mathrm{FB}}\rho_{\mathrm{B}}(\mathbf{r},t)$
for fermions and
$\lambda_{\mathrm{FB}}\rho_{\mathrm{F}}(\mathbf{r},t)$ for bosons.

The collective modes are small amplitude oscillations of the coupled
BEC and Fermi-liquid that propagate as a plane wave with a
wavevector $\mathbf{k}$. We assume $\mathbf{k}$ is significantly
shorter than the Fermi wavenumber. The phase
$\theta_{\mathrm{B}}(\mathbf{r},t)$ and density $\rho_{\mathrm{B}
}(\mathbf{r},t)$ of the BEC field $\left\langle \psi\left(
\mathbf{r} ,t\right)  \right\rangle =\sqrt{\rho_{\mathrm{B}}\left(
\mathbf{r},t\right) }e^{i\theta\left(  \mathbf{r},t\right)  }$ vary
as the real parts of $\theta(t)\exp(i\mathbf{k}\cdot\mathbf{r})$ and
$\eta(t)\rho_{\mathrm{B}} ^{0}\exp(i\mathbf{k}\cdot\mathbf{r})$,
with fluctuation amplitudes $\theta(t)\ll1$ and $\eta(t)\ll1$. The
Fermi-liquid oscillation can be described by a fluctuation of a
distribution function $n(\mathbf{r} ,\mathbf{p};t)$, with
$\mathbf{r}$ the position and $\mathbf{p}$\ the momentum. The
distribution function evolves according to the collisionless
transport equation \cite{PN},
\begin{equation}
\frac{\partial n}{\partial t}+\frac{\mathbf{p}}{m_{\mathrm{F}}}\cdot
\frac{\partial n}{\partial\mathbf{r}}-\overline{\nabla}\left(  \lambda
_{\mathrm{FB}}\rho_{\mathrm{B}}\right)  \cdot\frac{\partial n}{\partial
\mathbf{p}}=0,\label{eq:dn/dt}%
\end{equation}
where we have approximated the velocity of the dilute fermion gas as
$\mathbf{v}\simeq\mathbf{p}/m_{\mathrm{F}}$ and the force to be the
BEC-induced mean-field gradient, $F=-\overline{\nabla}\left[
\lambda _{\mathrm{FB}}\rho_{\mathrm{B}}\left(  \mathbf{r},t\right)
\right]  $. Assuming $n$ fluctuates around the equilibrium
distribution $n^{0}\left( \mathbf{p}\right)  $ by deforming the
local Fermi-surface, $n$ is linearized as
\begin{equation}
n\left(  \mathbf{r},\mathbf{p},t\right)  =n^{0}\left(  \mathbf{p}\right)
+u_{\mathbf{p}}\left(  t\right)  p_{\mathrm{F}}\delta\left(  \mathbf{p}%
-p_{\mathrm{\ F}}\right)  e^{i\mathbf{k}\cdot\mathbf{r}},
\end{equation}
where the second term is the fluctuation part $\delta
n_{\mathbf{p}}\left( \mathbf{r},t\right)  $, $p_{\mathrm{F}}$ is the
equilibrium Fermi-momentum, and $u_{\mathbf{p}}\left(  t\right)  $
is the amplitude of deformation. Then Eq.\ (\ref{eq:dn/dt}) becomes
\begin{equation}
\frac{\partial}{\partial t}u_{\mathbf{p}}\left(  t\right)  +i\frac
{\mathbf{k}\cdot\mathbf{p}}{m_{\mathrm{F}}}u_{\mathbf{p}}\left(
t\right)
=-i\frac{\mathbf{k}\cdot\mathbf{p}}{p_{\mathrm{F}}^{2}}\left(
\lambda_{\mathrm{FB}}\rho_{\mathrm{B}}^{0}\right)  \eta\left(
t\right)
.\label{eq:fermiondynamics}%
\end{equation}

Next, we describe the BEC dynamics. The hydrodynamic description of
BEC equations are \cite{PS}
\begin{equation}
\frac{\partial\rho_{\mathrm{B}}\left(  \mathbf{r},t\right)
}{\partial t}=-\overline{\nabla}\cdot\left(  \rho_{\mathrm{B}}\left(
\mathbf{r} ,t\right)
\frac{\hbar}{m_{\mathrm{B}}}\overline{\nabla}\theta_{\mathrm{B}
}\left(  \mathbf{r},t\right)  \right)  ,\label{eq:BECdensity}
\end{equation}%
\begin{align}
-\hbar\frac{\partial\theta_{\mathrm{B}}\left(  \mathbf{r},t\right)
}{\partial t} &
=\frac{-\hbar^{2}\overline{\nabla}^{2}\sqrt{\rho_{\mathrm{B}}\left(
\mathbf{r},t\right)  }}{2m_{\mathrm{B}}\sqrt{\rho_{\mathrm{B}}\left(
\mathbf{r},t\right)  }}\nonumber\\
&  +\lambda_{\mathrm{BB}}\rho_{\mathrm{B}}\left(  \mathbf{r},t\right)
+\lambda_{\mathrm{FB}}\rho_{\mathrm{F}}\left(  \mathbf{r},t\right)
,\label{eq:BECphase}%
\end{align}
where $\lambda_{\mathrm{BB}}$ is the boson-boson interaction
strength,
$\lambda_{\mathrm{BB}}=(4\pi\hbar^{2}a_{\mathrm{BB}})/m_{\mathrm{B}}$.
The coupling to the Fermion system arises from the mean-field
interaction term $\lambda_{\mathrm{FB}}\rho_{\mathrm{F}}\left(
\mathbf{r},t\right)
\approx\lambda_{\mathrm{FB}}\rho_{\mathrm{F}}^{0}+\lambda_{\mathrm{FB}}
(p_{F}^{3}/2\pi^{2})\left\langle u(t)\right\rangle \exp(i\mathbf{k}
\cdot\mathbf{r})$ with $\left\langle u(t)\right\rangle $ the angular
average, evaluated on the Fermi surface, $\left\langle
u(t)\right\rangle =(4\pi )^{-1}\int u_{\mathbf{p}}\left(  t\right)
d\Omega_{\mathbf{p}}$, $|{\mathbf{p}}|=p_{F}$.\textsl{\ }Linearizing
the fluctuation, combining Eqs.\ (\ref{eq:BECdensity}) and
(\ref{eq:BECphase}), then using the Bogoliubov dispersion
$\omega_{k}^{B}=kc\sqrt{1+(k\xi)^{2}}$, where $c$ is the single
BEC-sound velocity, $c=\left(
\lambda_{\mathrm{BB}}\rho_{\mathrm{B}} ^{0}/m_{\mathrm{B}}\right)
^{1/2}$, and $\xi$ is the BEC coherence length, $\xi=\left(
16\pi\rho_{\mathrm{B}}^{0}a_{\mathrm{BB}}\right)  ^{-1/2}$, we
obtain the collective BEC-dynamics,
\begin{equation}
\frac{\partial^{2}\eta(t)}{\partial t^{2}}=-\left(
\omega_{k}^{\mathrm{B} }\right)  ^{2}\eta(t)-3k^{2}c^{2}\left(
\frac{\lambda_{\mathrm{FB}}
\rho_{\mathrm{F}}^{0}}{\lambda_{\mathrm{BB}}\rho_{\mathrm{B}}^{0}}\right)
\left\langle u(t)\right\rangle .\label{eq:BECdynamics}
\end{equation}

Substituting Eq.\ (\ref{eq:fermiondynamics}) into Eq.\
(\ref{eq:BECdynamics}) and imposing harmonic time dependence,
$\eta(t)=\overline{\eta}\exp(-i\omega t)$, $u_{\mathbf{p}}\left(
t\right)  =\overline{u}_{\mathbf{p}}\exp(-i\omega
t)$ give the collective mode equation%
\begin{equation}
\frac{s^{2}\left(  v_{\mathrm{F}}/c\right)  ^{2}-\left(
1+q^{2}\right)  } {F}=\frac{1}{2}\int_{-1}^{1}\frac{x}{s-x}dx,
\label{eq: collective dynamics}
\end{equation}
where%
\begin{equation}
F=\left(  \frac{k_{\mathrm{F}}a_{\mathrm{FB}}}{2\pi}\right)  \left(
\frac{a_{\mathrm{FB}}}{a_{\mathrm{B}}}\right)  \left(
1+\frac{m_{\mathrm{F}} }{m_{ \mathrm{B}}}\right)  \left(
1+\frac{m_{\mathrm{B}}}{m_{\mathrm{F}} }\right)  \;,
\end{equation}
with $k_{\mathrm{F}}=p_{\mathrm{F}}/\hbar$, and where
$s\equiv\omega/(kv_{ \mathrm{F} })$ and $q\equiv k\xi$. The integral
on the right hand side stems from the angular average ($x$ is the
cosine of the angle between $\mathbf{k}$ and $\mathbf{p}$). Eq.\
(\ref{eq: collective dynamics}) is an s-wave zero-sound dispersion
with a Fermi-liquid parameter \cite{Baym} that has a resonant
denominator
\begin{equation}
F_{0}=F/\left[  s^{2}\left(  v_{\mathrm{F}}/c\right)  ^{2}-\left(
1+q^{2}\right)  \right]  .
\end{equation}

Eq.\ (\ref{eq: collective dynamics}) also reveals that the mode
structure depends on two relevant ratios: the velocity ratio
$c/v_{\mathrm{F}}$ and the interaction parameter, $F$. The velocity
ratio $c/v_{\mathrm{F}}$ provides the ratio of time scales. The
oscillation periods of a long wavelength BEC and a collective
fermion mode of wavenumber $k$ are $\sim(kc)^{-1}$ and $\sim(
kv_{\mathrm{F}})^{-1}$, respectively. They are also the times that
take the quantum liquids to respond to a density perturbation of
spatial variation on the length scale $k^{-1}$.
\begin{figure}[tbh]
\begin{center}
\includegraphics[width=3.1in]{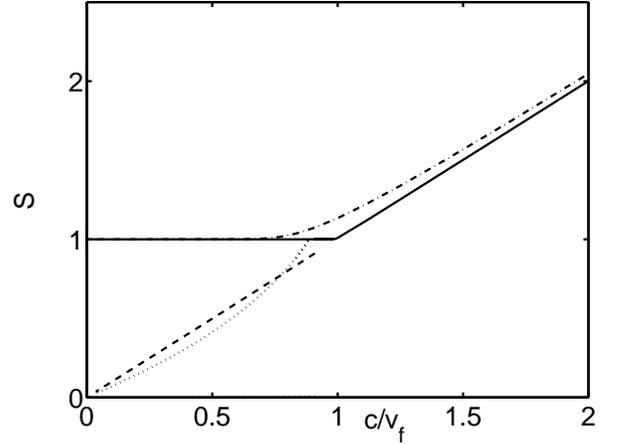}
\end{center}
\caption{The scaled velocity $s$ (in units of the Fermi-velocity)
for the collective damped and undamped modes as a function of
$c/v_{F}$ for $q=0$. solid : undamped mode for $F=0.01$,
dash-dotted: undamped mode for $F=0.5$, dashed: damped mode for
$F=0.01$, and dotted: damped mode for $F=0.5$}
\label{Fp01Fp5mode12}%
\end{figure}

In the long wavelength limit (i.e., $k\rightarrow0$), if the
BEC-response is significantly slower than the fermion frequency,
$c/v_{\mathrm{F}}\ll1$, the BEC fluctuation fails to follow the
fermion fluctuation and oscillates out-of-phase, giving an effective
$\emph{repulsive}$ fermion-fermion interaction as opposed to the
attractive one in the static limit. In this regime, the undamped
solution to Eq.\ (\ref{eq: collective dynamics}) describes a
collective excitation with the characteristics of the zero-sound
mode of a pure fermion gas with weak repulsive interactions. Its
velocity, $v $ is nearly equal to $v_{\mathrm{F}}$ (see Fig.\
\ref{Fp01Fp5mode12}). On the other hand, in the limit,
$c/v_{\mathrm{F}}\gg1$, the BEC is `fast' enough to be able to
follow fermion oscillations adiabatically, thus validating the
static description of BEC-mediated fermion-fermion interactions.
Since the effective interaction is attractive, the mode is not a
pure Fermi-liquid form of zero-sound, but the one that resembles a
pure BEC with $v\approx c$.

As a function of $c/v_{\mathrm{F}}$, the undamped collective
excitation crosses over from pure fermion-like zero sound to pure
BEC-like sound around $c/v_{\mathrm{F}}=1$, but the width of the
transition regime depends on $F$. Setting $s=1+\epsilon$,
$\epsilon\ll1$, the beginning of the transition can be estimated as
$\left(  1+F\left[  -1+\left(  1/2\right)  \ln\left(
\epsilon/2\right)  \right]  \right)  ^{-1/2}$. A crossover from pure
fermion-like zero-sound to a pure BEC-like sound mode is also
evident from the collective mode dispersion if
$c/v_{\mathrm{F}}\leq1$ \footnote{From $\xi
k_{\mathrm{F}}=(m_{\mathrm{F}}/2m_{\mathrm{B}})(v_{\mathrm{F}}/c)$,
we find that even if $c$ and $v_{\mathrm{F}}$ are comparable, the
inverse BEC coherence length can remain significantly smaller than
$p_{\mathrm{F}}$ provided the mixture consists of light boson atoms
immersed in a Fermi-sea of much heavier fermion atoms (e.g hydrogen
immersed in $^{40}$K, or $^{7}$Li mixed with $^{84}$Rb). In that
case, Eq.(\ref{eq: collective dynamics}) can describe excitations of
wavelength comparable to $\xi^{-1}$.}.
\begin{figure}[tbh]
\begin{center}
\includegraphics[width=3.1in]{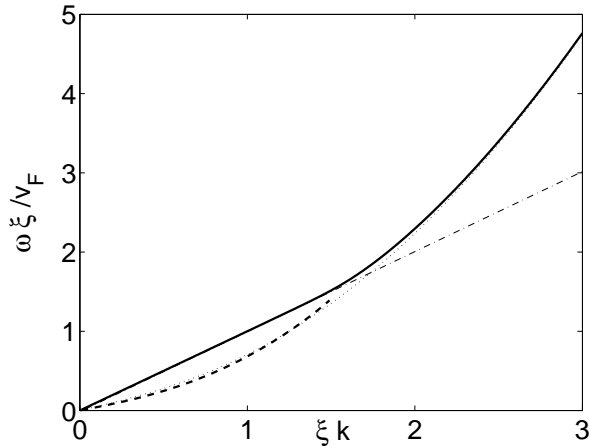}
\end{center}
\caption{Scaled dispersion relation for $F=0.5$ and $v_{F}/c=2$. The
dispersion changes from a linear pure fermion-like zero-sound
dispersion (dash-dotted line) to a pure BEC-like Bogoliubov
dispersion (dotted lime).} \label{dispFp5a2}
\end{figure}
Figure \ \ref{dispFp5a2} shows the avoided cross-over from a linear
to a Bogoliubov dispersion. In the cross-over region centered around
$k\simeq\xi^{-1}\sqrt{\left(  v_{\mathrm{F}}/c\right)  ^{2}-1}$. For
mixtures within this transition regime, the undamped collective mode
is a strong admixture with boson and fermion density fluctuations of
comparable amplitude. In the language of optics, the mixture changes
its index of refraction in the spatial region in which the trapped
mixture has a local velocity ratio within the transition range. This
shell can act as a boundary region from which a propagating
wavepacket of density fluctuations is reflected.

In a mixture with $c/v_{\mathrm{F}}<1$, the mediated interactions
remain weak so that the Bogoliubov mode of the single BEC appears as
a second mode with a dispersion that is slightly modified by the
fermion mediated boson-boson interactions. The zero-sound
deformation of the local Fermi surface creates a cloud of fermion
density fluctuations that accompanies (`dresses') the propagating
BEC-excitation. This collective excitation damps by Landau damping
and the complex solution $s=r-i\gamma$ with $0<r<1$, $\gamma>0$.
Finding this mode is rather delicate: A damped excitation,
$\exp\left[  -i\left( \omega-i\Gamma\right)  t\right]  $ where
$\Gamma=kv_{ \mathrm{F}}\gamma$, requires $r-i\gamma$ in the
left-hand side of Eq.\ (\ref{eq: collective dynamics}), while the
integral on the right-hand side, which describes a retarded mediated
interaction, requires the replacement $s\rightarrow r+i\gamma$. The
numerical evaluation shows that, for $c/v_{ \mathrm{F}}<1$ and
$F\ll1$, the damping rate can be well-approximated by the Fermi
golden rule for phonon annihilation accompanied by a fermion
particle-hole excitation: $\Gamma=ck\left(  \pi F/4\right)  \left(
c/v_{ \mathrm{F}}\right)  =\gamma\tau_{0}^{-1}$, where
$\tau_{0}=\left(  m_{ \mathrm{F}}/2m_{\mathrm{B}}\right)  \left(
ck_{\mathrm{F}}\right)  ^{-1}$ is the lifetime of the mode. In atom
trap experiments, $\tau_{0}$ can be order of milliseconds -- long
enough to measure the damping but short enough to keep a propagating
wavepacket within the spatial region where the system is
approximately homogeneous. Even for a rather large value of $F$,
such as $F=0.5$, $\Gamma$ obtained from the numerical evaluations
agree with the one using Fermi Golden rule with $30$\% error at
$c/v_{\mathrm{F}}\sim0.8$, where the damped zero-sound mode
vanishes.

The damped mode dispersion relation in Fig.\ \ref{dispFp5a2}
terminates at $k$ before entering the crossover region. This
termination point is reminiscent of the $^{4}$He-mode for which the
termination is connected with the decay into quasi-particle pairs
\cite{Landau}.

While the imaginary part of the dispersion relation merely describes
the damping of the collective mode and not the instability of the
homogeneous mixture as claimed by Ref.\ \cite{PZWM02}, Eq.\
(\ref{eq: collective dynamics}) does give insight into the stability
issue. When the interaction parameter $F$\ exceeds unity, the damped
mode $s$ takes on a purely imaginary value, $s=i\gamma$, on account
of which the right-hand side of Eq.\ (\ref{eq: collective dynamics})
reduces to $\gamma\arctan(\gamma^{-1})-1$ \cite{Inguscio} and which
does signal the instability of the homogeneous mixture. A degenerate
fermion system of $\rho_{\mathrm{\ F}}^{0}
>\rho_{\mathrm{F},\mathrm{crit}}$,
\begin{equation}
\rho_{\mathrm{F},\mathrm{crit}}=\frac{4\pi}{3a_{\mathrm{FB}}^{3}}\left[
\frac{a_{\mathrm{BB}}/a_{\mathrm{FB}}}{(1+m_{\mathrm{B}}/m_{\mathrm{F}
})(1+m_{\mathrm{F}}/m_{\mathrm{B}})}\right]  ^{3},
\end{equation}
is immiscible to a BEC and an initially homogeneous mixture
spontaneously separates into either regions of pure fermion and pure
BEC systems or regions of pure fermion and mixed fermion-BEC phases
\cite{Viverit}. The experimentalist can realize the separation by
tuning an external magnetic field near the zero-point of the
boson-boson scattering length $a_{\mathrm{\ BB}}$ in a Feshbach
resonance \cite{Molmer}. Eq.\ (\ref{eq: collective dynamics}) also
provides the dynamics of the onset of the instability. In response
to a sudden change of $a_{\mathrm{BB}}$ in an initially homogeneous
gas mixture, the quantum gases gather into `clumps' of single phase
matter before the clumps congregate into larger regions of single
phase matter.

The clumping is initiated by the exponential growth of the unstable
collective mode eigenvectors of wavenumber $k=q\xi^{-1}$ and grows
at a rate $\gamma q\tau_{0}^{-1}$. The fastest growing eigenvectors
of wavenumber $q^{\prime}$ dominate the dynamics and determine the
size of the clumps as well as the rate of formation \cite{BECphase}.
By phase separating a BEC-fermion mixture in a cigar-shaped trap,
the experimentalist can prevent the single phase clumps to move past
each other and measure the size of the clumps frozen in place, as
demonstrated in the phase separation of BEC's \cite{Ketterleps}. If
the instability is sufficiently weak, $F-1\ll1$, the right-hand side
of Eq.\ (\ref{eq: collective dynamics}), is well approximated by
$-\gamma (\pi/2)-1$ and the unstable eigenvectors $q\in\left(
0,\sqrt{F-1}\right)  $ grow at at a rate $\gamma q\tau_{0}^{-1}$
with $\gamma(q)=(\pi F/4)+\sqrt{(\pi
F/4)^{2}-(v_{\mathrm{F}}/c)^{2}\left[  1+q^{2}-F\right]  }$, giving
a dominant wavenumber $q^{\prime}\xi^{-1}$ with
\begin{align}
q^{\prime}{}^{2}  &  =\frac{1}{128}\left(  64(F-1)+F\pi\left(
c/v_{\mathrm{F}
}\right)  ^{2}\times\right. \nonumber\\
&  \;\left.  \left[  3F\pi-\sqrt{(3F\pi)^{2}+128(v_{\mathrm{F}}/c)^{2}\left(
F-1\right)  }\right]  \right)  \;.
\end{align}
We expect the average clump size to be $\sim\xi(2\pi/q^{\prime})$
and the clumps to form on a time $\sim\tau_{0}\left(  \gamma
q^{\prime}\right)  ^{-1}$.

Finally, we remark that previously developed cold atom
BEC-techniques are ideally suited to probe the collective modes of
fermion-boson mixtures. A focused laser beam or crossed beams in a
two-photon Bragg-scattering setup can create a density variation
near the middle of the trap, the subsequent propagation of which can
be imaged optically or by a time-of-flight measurement. The dual
mode structure of well separated group velocities would manifest
itself by spatially separating the initial wave into two wavepackets
with the faster one moving at a speed $v_{\mathrm{F}}$ instead of
the speed of ordinary sound ($v_{\mathrm{F}}/\sqrt{3}$ in dilute
fermions). The $40$\% difference in velocity, in stark contrast to
the few percent difference in the strongly interacting
$^{3}$He--fluid, can be easily measured. The slower wavepacket moves
at a speed closer to $c$ and is damped. Furthermore, if the
spatially varying equilibrium fermion and BEC-densities in the atom
trap reach values where the local Fermi and BEC sound velocities are
equal, we expect that this region can act as a boundary from which
the faster wavepacket is reflected. The low energy collective modes
of the finite size system \cite{Capuzzi} can similarly be located
inside or outside the boundary.

In conclusion, we have shown that a quantum degenerate gas mixture
of a BEC and single component fermion gas exhibits a surprisingly
rich collective mode structure that is closely related to the
phenomenon of mediated interactions. Even though the boson mediated
fermion-fermion interactions are attractive in the static limit, the
mixture can support a long lived zero-sound mode that resembles that
of a pure Fermi-liquid in the resonant dynamics of the
phonon-mediation if $F<1$, since the effective interaction is
repulsive. As a function of the velocity ratio $c/v_{\mathrm{F}}$,
the undamped mode undergoes an avoided crossing near $c\sim
v_{\mathrm{F}}$ and then, becomes a single BEC-sound like mode as
$c/v_{\mathrm{F}}$ increases. A second zero-sound mode appears in
the slow BEC regime $c/v_{\mathrm{F}}<1$ . This mode undergoes
Landau damping and terminates. If $F>1$, the homogeneous mixture
becomes unstable and undergoes phase separation for which we derive
the relevant time and length scales.

D.H.S.'s work is supported by the NSF through a grant for the
Institute for Theoretical Atomic, Molecular and Optical Physics at
Harvard University and Smithsonian Astrophysical Observatory. D.H.S.
and E.T. are grateful to KITP, University of California at Santa
Barbara, for providing a stimulating environment during part of our
collaboration.


\begin{thebibliography}{99}                                                                                               %


\bibitem {GrimJinKetSalThom}C. Chin \textit{et al.}, Science, \textbf{305}, 1128
(2004); C. A. Regal, M. Greiner, and D. S. Jin, Phys. Rev. Lett.
\textbf{92}, 040403 (2004); M. W. Zwierlein et al., Phys. Rev. Lett.
\textbf{92}, 120403 (2004); T. Bourdel \textit{et al.}, Phys. Rev.
Lett. \textbf{93} 050401 (2004); K. M. O'Hara \textit{et al.},
Science \textbf{298}, 2179 (2002); R. Hulet, presentation in the
Quantum Gas Conference at the Kavli Institute for Theoretical
Physics, Santa Barbara, CA, May 10-14 (2004).

\bibitem {EXPT0102}A. G. Truscott \textit{et al.}, Science \textbf{291}, 2570 (2001); F.
Schreck \textit{et al.}, Phys. Rev. Lett. \textbf{87}, 080403
(2001); Z. Hadzibabic \textit{et al.}, Phys. Rev. Lett. \textbf{88},
160401 (2002); G. Roati \textit{et al.}, Phys. Rev. Lett.
\textbf{89}, 150403 (2002).

\bibitem {Eddy}E. Timmermans, Phys. Scripta, \textbf{T 110}, 302 (2004).

\bibitem {Baym}G. Baym and C. Pethick, \textit{Landau Fermi-liquid theory}
(Wiley-Interscience, New York, 1991).

\bibitem {KS93}E. Krotscheck and M. Saarela, Phys. Rep. \textbf{232}, 1 (1993).

\bibitem {BBP66}J. Bardeen, G. Baym, and D. Pines, Phys. Rev. Lett.\textbf{
17} (7), 372 (1966).

\bibitem {Pines}D. Pines, \textit{Elementary Excitations in Solids}, (Perseus
Books Group, New York, 1963).

\bibitem {Stoof}M. J. Bijlsma, B. A. Heringa, and H. T. C. Stoof, Phys. Rev.
A., \textbf{61}, 053601 (2000).

\bibitem {PZWM02}Han Pu, Weiping Zhang, Martin Wilkens, and Pierre Meystre,
Phys. Rev. Lett. \textbf{88}, 070408 (2001).

\bibitem {PN}D. Pine and P. Nozieres, \textit{Theory of Quantum Liquids}
(Addison Wesley Publishing Company, Reading, 1994).

\bibitem {PS}C.J. Pethick and H. Smith, \textit{Bose-Einstein Condensation in
Dilute Gases} (Cambridge University Press, Cambridge, 2001).

\bibitem {Landau}E.M. Lifshitz and L. P. Pitaevskii, \textit{Statistical
Physics, Part 2} (Butterworth-Heineman, Oxford, 1991).

\bibitem {Inguscio}R. Roth, Phys. Rev. A \textbf{66}, 013614 (2002); G.
Modugno, G. Roati, F. Riboli, F. Ferlaino, R. J. Brecha, and M. Inguscio,
Science, \textbf{297} 2240 (2002).

\bibitem {Viverit}L. Viverit, C. J. Pethick, and H. Smith, Phys. Rev. A,
\textbf{61}, 053605 (2000).

\bibitem {Molmer}K. Molmer, Phys. Rev. Lett. \textbf{80}, 1804 (1998).

\bibitem {BECphase}E. Timmermans, Phys. Rev. Lett. \textbf{81}, 5718 (1998).

\bibitem {Ketterleps}J. Stenger \textit{et al.}, Nature, \textbf{396} 345 (1998).

\bibitem {Capuzzi}M. Amoruso \textit{et al.}, Eur.
Phys. J. D, \textbf{11}, 335 (2000); P. Capuzzi and E. S. Hernandez,
Phys. Rev. A \textbf{64}, 043607 (2001); T. Miyakawa, T. Suzuki, and
H. Yabu, Phys. Rev. A \textbf{62}, 063613 (2000).

\end{thebibliography}
\end{document}